\documentclass[conference]{IEEEtran}

\usepackage{graphicx}

\usepackage{subfigure}
\usepackage{url}
\usepackage{longtable}
\usepackage{xtab,afterpage}
\usepackage[nolist,nohyperlinks]{acronym}
\usepackage{framed}
\usepackage{etoolbox}
\usepackage{paralist}
\usepackage{multirow,booktabs}
\usepackage{amsmath}
\usepackage{csvsimple}
\usepackage{tikz}
\usepackage{amssymb}
\usetikzlibrary{patterns,shapes,arrows,calc,shadows}
\usepackage{pgfplots}
\usepackage{mathtools}

\usepackage{tikz}
\usetikzlibrary{arrows,decorations.pathmorphing,backgrounds,positioning,fit,petri}
\usetikzlibrary{shapes.misc}
\tikzset{cross/.style={cross out, draw=black, minimum size=2*(#1-\pgflinewidth), inner sep=0pt, outer sep=0pt},
	cross/.default={1pt}}
\usetikzlibrary{patterns}

\usepackage[noadjust]{cite}

\usepackage{pgfplots}
\usepackage{bm}
\usepackage{amssymb}
\usepackage{mathtools}
\usepackage{standalone}
\usepgfplotslibrary{groupplots}
\usepackage{caption}

\usetikzlibrary{arrows,backgrounds,positioning,fit}

\usetikzlibrary{arrows,shadows}
\usepackage{pgf-umlsd}
\usepackage{algorithm}
\usepackage[noend]{algpseudocode}
\algnewcommand{\Initialize}[1]{%
	\State \textbf{Initialize:}
	\Statex \hspace*{\algorithmicindent}\parbox[t]{.8\linewidth}{\raggedright #1}
}

\usepackage{textcomp}

\graphicspath{{images/}{images/}}

\newtoggle{authorcomment}
\toggletrue{authorcomment}

\usepackage[font=small]{caption}

\usepackage{authblk}

\usepackage{hyperref} 
\hypersetup{backref,hidelinks,colorlinks=false,breaklinks=true,bookmarksopen=false,pdftitle={Title},pdfauthor={Author}}

\usepackage{tikz}
\usetikzlibrary{automata, positioning}

\usepackage{verbatim} 
\usepackage{booktabs}
\usepackage{color, colortbl}
\definecolor{LightCyan}{rgb}{0.88,1,1}
\definecolor{Gray}{gray}{0.85}

\begin{document}

\title{Using Deep Q-learning To Prolong the Lifetime of Correlated Internet of Things Devices}

\author{Jernej Hribar$^*$, Andrei Marinescu$^*$, George A. Ropokis$\dag$, and Luiz A. DaSilva$^*$ \\ $^*$CONNECT- Research Centre at Trinity College Dublin, Dublin 2, Ireland \\ 
$\dag$ CentraleSup\'elec/IETR, CentraleSup\'elec Campus de Rennes, 35510
Cesson Sevigne, France\\ \textit {E-mail: jhribar@tcd.ie.}}



\maketitle

\begin{abstract}
\thispagestyle{empty}
Battery-powered sensors deployed in the \ac{iot} require energy-efficient solutions to prolong their lifetime. 
When these sensors observe a physical phenomenon distributed in space and evolving in time, the collected observations are expected to be correlated. 
We take advantage of the exhibited correlation and propose an updating mechanism that employs deep Q-learning. Our mechanism is capable of determining the frequency with which sensors should transmit their updates while taking into the consideration an ever-changing environment.
We evaluate our solution using observations obtained in a real deployment, and show that our proposed mechanism is capable of significantly extending battery-powered sensors' lifetime without compromising the accuracy of the observations provided to the \ac{iot} service.
\end{abstract}
\acresetall


\section{Introduction}
\label{sec:intro}

Over the next few years, billions of devices are expected to be deployed in the \ac{iot} network \cite{ericsson2018report}, many of which will be low-cost sensors powered by non-rechargeable batteries. These battery-powered sensors will provide vital information regarding the environment to numerous services, e.g. smart farming \cite{wolfert2017big}, autonomous vehicles \cite{gerla2014internet}, air pollution monitoring \cite{al2010mobile}, etc. 
Providing accurate and up-to-date information to services while keeping the battery-powered sensors functional as long as possible is one of the primary challenges in the \ac{iot}. 

It is possible to take advantage of the correlation exhibited in measurements when multiple sensors measure the same physical phenomenon occurring in the environment.  In our work, we utilise the correlation in the data collected by multiple sensors to prolong battery-powered sensors' lifetime.
In particular, we have designed an energy-efficient updating mechanism capable of taking advantage of correlation exhibited in the collected information by applying a \ac{rl} \cite{sutton1998reinforcement} technique.

Most existing research that applies \ac{rl} in the context of \ac{iot} has focused on exploiting devices' behaviour in the physical layer to improve their energy performance. For example, in \cite{li2018q} authors used Q-learning to enhance the spectrum utilisation of industrial \ac{iot} devices. They demonstrated that devices are capable of learning a channel selection policy to avoid collisions, thus improving their energy efficiency. Similarly, in \cite{chu2012aloha}, the authors applied Q-learning to improve device access to the channel, to avoid collisions. The use of deep \ac{rl} was investigated in \cite{mohammadi2018semisupervised}, where authors relied on Bluetooth signal strength to improve indoor users' location estimation. In contrast to the works mentioned above and those described in \cite{alsheikh2014machine} focusing mostly on the physical layer, we learn from information collected. Applying \ac{rl} to learn from the content of information collected to prolong the sensors' lifetime has not been proposed before, to the best of our knowledge. 




In this paper, we propose an updating mechanism capable of learning the frequency of updates, i.e., how often an \ac{iot} sensor should transmit updated readings. Our approach prolongs battery-powered sensor's lifetime by leveraging correlation exhibited in observations collected, without hindering the accuracy and timeliness of the information provided to services relying on these observations. We define the decision-making problem 
that our proposed mechanism is capable of solving in Section \ref{sec:sensing_system}. To solve the proposed problem using \ac{dqn} \cite{mnih2013playing}, we describe the system dynamics using states, actions, and rewards from an \ac{rl} perspective (Section \ref{sec:rl_approach} B). We also describe the overall mechanism in the form of a block diagram (Section \ref{sec:rl_approach} C). We evaluate the proposed mechanism using data obtained from a real deployment and show that the system is capable of prolonging the minimum expected sensor lifetime by over two and a half times (Section \ref{sec:eval}). Finally, we discuss open issues and our future work (Section \ref{sec:conclusion}).

\section{Problem formulation}
\label{sec:sensing_system}
\begin{figure}
	\centering
	\includegraphics[width=3.5in]{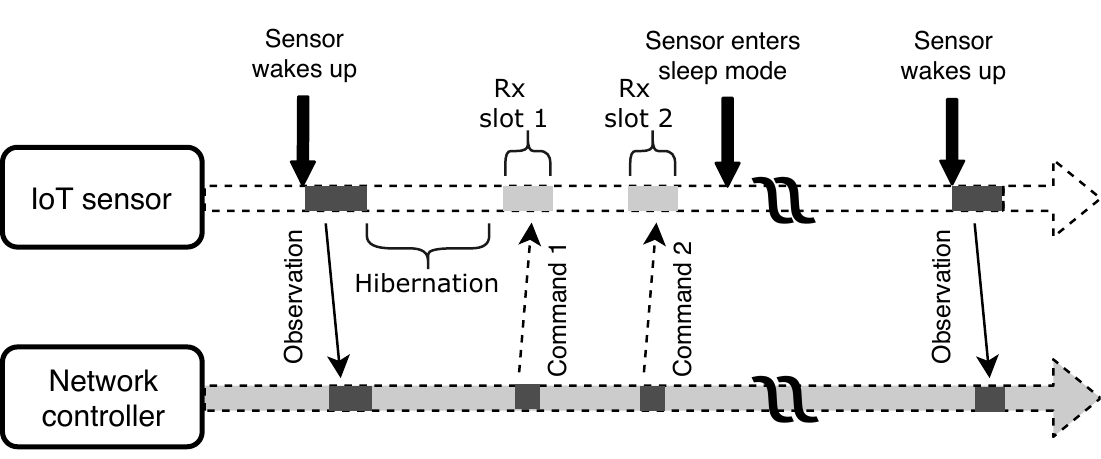}
	\caption{Message sequence of an \ac{iot} sensor using \ac{lorawan}.}
	\label{fig:Lora_message_sequence}
\end{figure}

In our work, we focus on inexpensive battery-powered sensors transmitting observations. These sensors are constrained in terms of available computing power, communication capabilities, and energy. Furthermore, such sensors rely on the use of sleep mode to preserve energy. When a sensor is in sleep mode, the rest of the network cannot communicate with it. Consequently, the network controller, responsible for collecting observations, has to inform each sensor, while the sensor is still in active mode, when it should wake up again and transmit the next observation. The low power \ac{iot} sensor is usually in active mode only after it has transmitted. For example, a sensor using \ac{lorawan} class A radio, will listen for two short time-windows after it has transmitted, as illustrated in the \ac{lorawan} message sequence in Fig. \ref{fig:Lora_message_sequence} \cite{loradatasheet}. 

\begin{figure}
	\centering
	\includegraphics[width=3.5in]{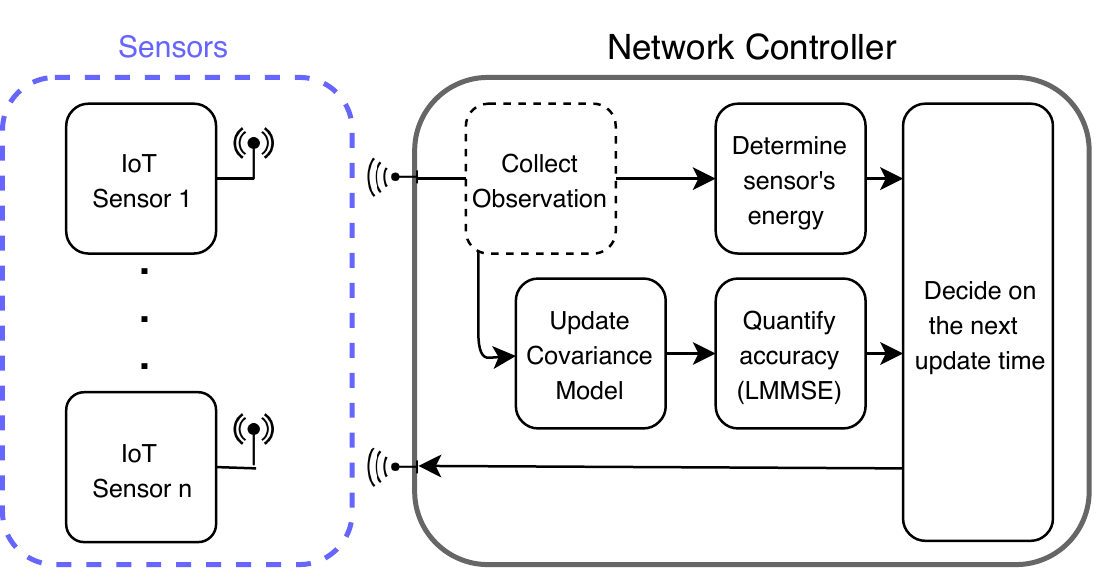}
	\caption{A high-level overview of the decision-making process in a network controller managing $N$ IoT sensors.}
	\label{fig:problem_diagram}
\end{figure}

The network controller has two objectives when setting the transmission times for sensors. The first objective is to satisfy the accuracy goals set by services for the collected observations. The second objective is to prolong the lifetime of battery powered sensors. Both objectives are achievable when sensors gather correlated information, as we demonstrated in \cite{hribar2018using}; our previous work, however, we did not propose a procedure for the controller to establish the desired frequency of updates. The network controller decides on the sensors' next update time by evaluating the accuracy of collected observation and the sensors' available energy. We summarize the decision-making process by the network controller in Fig. \ref{fig:problem_diagram}. In what follows, we present our methodology for modelling the observations' accuracy as well as the network controller decision process.

\subsection{Quantifying observations' accuracy: LMMSE}

We consider a network observing a phenomenon distributed over a geographical area and evolving in time $t$. The network employs $N$ sensors deployed at positions $\bm{x}_{n}, n=1, \ldots, N$. We use notation $Z\left(\bm{x}_n, t\right)$ to denote the value of the observed physical process at time instance $t$, at location $\bm{x}_{n}$. We can write collected observations into a vector $\mathbf{y} = \left[y_1, \ldots, y_{N} \right]^T$ with $y_n = Z \left(\bm{x}_{n}, t_{n} \right)$ where $t_n \in \left[0, t\right]$ is the latest time at which sensor $n$ has reported an observation. Then, using a simple \ac{lmmse} estimator, we can approximate the measurement at position $\bm{x}$ at time instance $t$, as:


\begin{equation}\label{eq:estimation}
\hat{Z}(\bm{x},t) = a_0 + \sum_{n=1}^{N} a_n y_n,
\end{equation}



\noindent where $a_n, n=0, \ldots , N$ are \ac{lmmse} estimator weights. We obtain the \ac{lmmse} estimator weight vector $\mathbf{a} = \left[a_0, \ldots, a_{N} \right]^T$ as follows:
\begin{equation}\label{eq:lmmse}
\mathbf{a} = (\bm{C_{YY}})^{-1} \bm{C_{YZ}}.
\end{equation}

\noindent The $\bm{C_{YY}}$, $\bm{C_{YZ}}$ are covariance matrices:
\begin{equation}\label{eq:covariance_matrix}
\bm{C_{YY}}=\begin{bmatrix}
\sigma \rho_{y_{1}y_{1}} \hdots \sigma  \rho_{y_1y_{N}}  \\
\vdots \phantom{ABCDE}  \vdots \\
\sigma \rho_{y_{N}y_{1}} \hdots \sigma \rho_{y_{N}y_{N}} \\
\end{bmatrix}
; \bm{C_{YZ}}=
\begin{bmatrix}
\sigma \rho_{y_{1}z} \\
\vdots \\
\sigma \rho_{y_{N}z} \\
\end{bmatrix};
\end{equation}


\noindent in which $\sigma$ represents standard deviation and $\rho$ represents covariance. Covariance describes how correlated the two observations are in time and space. For example, when temperature changes at one location, the covariance enables us to calculate the probability that the temperature has also changed by a certain value at another distant location. 
In our model, $\bm{C_{YY}}$ captures the covariance between observations at different locations, taken at the same time, and $\bm{C_{YZ}}$ captures the covariance between observations taken at the same location, at different times.
In this work, we adopt the separable covariance model defined in \cite{cressie1999classes} that allow us to express the correlation between two observations with time difference $\Delta_i(t)$ at locations at a distance $r_i$ as:

\begin{equation}\label{eq:correlation_II}
\rho_i(r_i,t) =   exp(-\theta_2(t) r_i -\theta_1(t) \Delta_i(t)).
\end{equation}


\noindent Note that $\theta_1(t)$ and $\theta_2(t)$ are scaling parameters of time and space, respectively. Both parameters change over time and are extracted from the obtained observations. In our work, we follow a scaling extraction method with Pearson's correlation coefficient formula for samples, as described in \cite{gneiting2002nonseparable}.

Using Eq. \eqref{eq:estimation} and Eq. \eqref{eq:lmmse} we can obtain estimates for the observed phenomenon, at the point $\bm{x}_n$, even at time instances in which the $n$-th sensor is in sleep mode. However, the system requires a way to evaluate how accurate these estimations are. For that purpose, we use the \ac{mse}:

\begin{equation}\label{eq:system_estimatio_error_n}
\varepsilon \big (\bm{x},t| \theta_1(t), \theta_2(t) \big ) = \sigma^2- \bm{C_{ZY}}\mathbf{a} ,
\end{equation}

\noindent where $\bm{C_{ZY}}$ is the transpose of $\bm{C_{YZ}}$ defined above.

By assessing the quality of estimates in the absence of fresh observations, the network controller can set the update times in such a way to ensure accurate collection of observations. However, the determined update time might not result in accurate estimation of the observed phenomenon due to changes in covariance model scaling parameters. The network controller should be able to anticipate such changes and act before they happen, i.e., while the sensor is still active. Additionally, the controller should be aware of sensors' available energy when deciding on sensors' next update time.

\subsection{Prolonging sensors' lifetime in a dynamic environment}
\label{sec:optimization}
The sensors' lifetime depends on the frequency of transmitted observations and on the continuous power consumption, i.e., the minimal power sensors always require to function. The sensor's lifetime can be simply modeled as in \cite{chen2005lifetime}: 


\begin{equation}\label{eq:lifetime}
\mathbb{E}[ \mathcal{L} ] = \frac{{E}_0}{P_c + \frac{\mathbb{E}[E_{tr}]}{T}},
\end{equation} 

\noindent where $P_c$ represents the continuous power consumption, $T$ denotes the time between updates, $\mathbb{E}[E_{tr}]$ represents the expected energy required to acquire and transmit the observation, and ${E}_0$ represents the sensors' starting energy. 


The network controller seeks to prolong the lifetime of battery-powered sensors by maximising
the time between two consecutive updates by a sensor, while keeping the accuracy of collected observations at every location of interest within the pre-specified boundary. In a real deployment, services dictate which locations are of interest. In this paper, we define every sensor location, i.e., $\bm{x}_{n}$, as a location of interest. The decision is also based the on the sensors' available energy, which the network controller can determine from the sensor's reported power supply measurement.
The network controller compares each sensors' energy with the energy available to other sensors and decides on the update time accordingly. Ideally, the system will set a higher update rate for a sensor with more available energy, to provide a longer lifetime to those sensors with less available energy. 

\vspace{8pt}


In our system, the \ac{mse} continuously varies, as 
every received observation can change the covariance model scaling parameters. These changes are application-specific: for example, in a smart factory, environmental changes are very frequent due to many factory processes simultaneously impacting the observed environment. In contrast, the changes in a smart farming scenario tend to be much more gradual. However, regardless of the frequency of changes, the system requires a means to adapt sensors' update time to the ever-changing environment. To that end, we propose for the network controller to employ \ac{rl} to decide on the next update time. Using \ac{rl}, the network controller can find a long-term updating policy to collect accurate observations and prolong battery-powered sensors' lifetime.

\section{Reinforcement Learning Approach}
\label{sec:rl_approach}
\ac{rl} allows an agent to learn its optimal behaviour, i.e., a set of actions to take in every state, solely from interactions with the environment. In our case, the agent is the network controller. The goal of learning is to determine the update time with which each sensor should transmit its observations.
Our agent observes the environment through the value of the \ac{mse} in the latest observation of the physical phenomenon reported by each sensor, as well as the remaining energy available to each sensor. 
The \ac{mse} is a non-stationary process, changing with every new observation, and the system exhibits non-Markovian properties. However, \ac{rl} has been proven to be applicable even when the system is non-Markovian \cite{jaakkola1995reinforcement}. In such a case, using an \ac{ann} enables the agent to reconstruct the hidden part of environment in the neural network. We implement the updating mechanism, i.e., the network controller decision process of when a sensor should transmit its next observation, using deep Q-learning \cite{mnih2013playing}.


\subsection{Q-learning}
The Q in Q-learning \cite{watkins1992q} stands for the quality of an action in a given state. The learning agent should take the action with the highest Q-value, unless the algorithm decides to explore. The agent learns the best action possible by updating the Q-value every time it takes an action and observes a reward. When the agent takes enough actions in every state of the environment, it can correctly approximate the real action values, i.e., Q-values, associated with every state. With Q-values determined, the agent can choose the optimal action in every state. A Q-value is calculated as follows:

\small
\begin{equation}\label{eq:q_value}
Q^{new}(s,a) \leftarrow 
Q(s,a) + \alpha \bigg ( R(s') + \gamma \max_{a'} Q(s',a')  - Q(s,a) \bigg )
\end{equation}
\normalsize

\noindent where $Q(s, a)$ is the previous Q-value, $\alpha$ is the learning rate, $\gamma$ is the discount factor, and $R$ is the reward observed in the new state $s'$ after taking action $a$ in state $s$. The $\max_{a'} Q(s',a')$ stands for an estimate of the optimal future value an agent can acquire from the next state $s'$.

The role of the agent, i.e., the network controller, is to determine the sensor state and to select the action with the highest Q-value. Every time a sensor transmits an observation, the network controller will respond by instructing the sensor for how long it should enter sleep mode, i.e., set its next update time. Next, we define states, actions, and reward functions, i.e., a tuple in $\langle \mathcal{S, A, R} \rangle$ that enables the network controller to determine sensors' optimal update times. 

\subsection{States, Actions, and Rewards}

Our \textit{state} space, $\mathcal{S}$, captures three critical aspects of the decision-making process in the network controller: a sensor's current update time, its available energy, and the estimation error value. Whenever a sensor transmits an observation, the network controller stores the information regarding sensor's update time, i.e., $T$, available energy, i.e., $E$, and value of average \ac{mse} since the last transmission, i.e., $\overline{\varepsilon}$. The learning agent then uses those values to represent each sensor state. The agent requires information regarding sensors' update times and \ac{mse} values to reconstruct the hidden part of the environment, while the energy levels enable the agent to ascertain which sensor in the network needs to save the energy the most. The total number of states is:

\begin{equation}\label{eq:state_space}
|\mathcal{S}| = (TE\overline{\varepsilon})^{N},
\end{equation}



\noindent with $N$ representing the number of sensors under the agent's control. 
Using \ac{ann} can efficiently solve problems associated with a sizable non-linear state space \cite{mnih2013playing}.

In contrast to the state space, we limit \textit{actions}, i.e., the cardinality of set $\mathcal{A}$, to five. Actions enable an agent to learn the best update time by decreasing or increasing the update time in time-steps. Furthermore, the agent can select to make a large or small step to adapt more quickly or more gradually. We denote an action to increase the update time for one time-step $U_{incr1}$ and for ten $U_{incr10}$. With $U_{dec1}$ and $U_{dec10}$ we denote decrease of update time for one and ten time-steps, respectively. If the agent decides to maintain the current update time unchanged, it selects action $U_{cons}$.
As we show in the next section, the system can using selected action space adapt to any changes in the environment promptly.
Additionally, limited action spaces prevent rapid fluctuations in the system when the learning algorithm decides to explore. 

We designed the \textit{reward function} to aid the agent to quickly adapt to the changing environment.
To achieve both network controller objectives, i.e., collecting accurate observations and prolonging sensors' lifetime, we split the reward function into two parts as follows: 

\begin{equation}
\mathcal{R}(s) = \phi_{acc}  \mathcal{R}_{acc}(s) +  \phi_{en}  \mathcal{R}_{en}(s).
\end{equation}

\noindent $\mathcal{R}_{acc}(s)$ rewards  accurate collection of observations and $\mathcal{R}_{en}(s)$ rewards energy conservation. We weigh each part of the reward function, with $\phi_{acc}$ for accuracy, and $\phi_{en}$ for energy. Our learning agent receives the reward after taking a decision for the update time and receiving the next update from the sensor. We define this learning cycle as one episode, i.e., our agent receives the reward after the end of each episode.


The accuracy reward depends on whether the set accuracy boundary, i.e., $\varepsilon_{tar}$, was satisfied in the episode. We compare the average \ac{mse} in an episode, i.e. $\overline{\varepsilon}(t)$, to the target \ac{mse}. The accuracy reward is as follows:

\begin{equation}
\mathcal{R}_{acc}(s) =  \begin{cases} 
\operatorname{ when }\phantom{A} \overline{\varepsilon}(t) \leq \varepsilon_{tar} \phantom{,}\operatorname{ :}\\
\phantom{-} 1 + \big (\frac{4\varepsilon_{tar}}{5} - \overline{\varepsilon}(t) \big)^3 T_{\Delta} + \frac{T_{\Delta}}{100}, \phantom{A} \operatorname{ if } \phantom{,} T_{\Delta}> 0 \\
\qquad \phantom{AA)} \qquad \phantom{-} \frac{3}{4}, \phantom{AAAAAAAA,}  \operatorname{ if } \phantom{,} T_{\Delta} = 0\\
\phantom{-A} -1 + T_{\Delta} \big(\varepsilon_{tar} - \overline{\varepsilon}(t) \big), \qquad \phantom{j} \operatorname{ if } \phantom{,} T_{\Delta} < 0\\
\operatorname{ when }\phantom{A} \overline{\varepsilon}(t) > \varepsilon_{tar} \phantom{,}\operatorname{ :}\\
\phantom{-A} -1 + T_{\Delta} \big(\varepsilon_{tar} - \overline{\varepsilon}(t) \big), \qquad \phantom{j} \operatorname{ if } \phantom{a} T_{\Delta} > 0\\
\qquad \phantom{AAA} \qquad - \frac{3}{4}, \phantom{AAAAAAA)}  \operatorname{ if } \phantom{,} T_{\Delta} = 0\\
\phantom{-,} 1 + \big (\frac{5\varepsilon_{tar}}{4} - \overline{\varepsilon}(t) \big)^3 T_{\Delta} - \frac{T_{\Delta}}{100}, \phantom{A} \operatorname{ if } \phantom{,} T_{\Delta} < 0 \\
\end{cases}
\end{equation}

\noindent where $T_{\Delta}$ represents the change in update time. When $\overline{\varepsilon}$ is below $\varepsilon_{tar}$, the agent receives a positive reward if it has increased the update time or kept it as it was. To provide an agent with a higher reward when the $\overline{\varepsilon}$ is very low and the selected action was to increase the update time for a few time-steps, we adopt a polynomial function (of order three) on the difference between the current $\overline{\varepsilon}$ and $\varepsilon_{tar}$. The same reward function will give a higher reward when the $\overline{\varepsilon}$ is close to $\varepsilon_{tar}$ and the change in update time is smaller. Such an approach is required to avoid overshooting the set accuracy boundary. Additionally, when the $\overline{\varepsilon}$ approaches the $\varepsilon_{tar}$ the reward for taking an action to decrease the update time slowly increases. We set the opposite reward function behaviour when the $\overline{\varepsilon}$ is above set $\varepsilon_{tar}$.


Our energy reward function depends on the change in update time and how a sensor's available energy compares to the average sensor's energy in the network. We write the energy reward as follows:
 
\begin{equation}
\mathcal{R}_{en}(s) =  \begin{cases}  \phantom{-A} \frac{2(E_{avg} - E)}{E_{avg}}, \qquad \qquad  \text{ if } \phantom{,} T_{\Delta}> 0 \\
\phantom{-AA} \frac{E_{avg} - E}{E_{avg}}, \qquad \qquad \phantom{,} \text{ if } \phantom{,} T_{\Delta}= 0 \\
\phantom{-A} \frac{2(E - E_{avg})}{E_{avg}}, \qquad  \qquad \text{ if } \phantom{,} T_{\Delta}< 0 \\
\end{cases},
\end{equation}

\noindent where $E$ is the sensor's available energy and $E_{avg}$ is the average energy available among all sensors in the network. 
	
\subsection{Proposed learning scheme}

\begin{figure}
	\centering
	\includegraphics[width=3.5in]{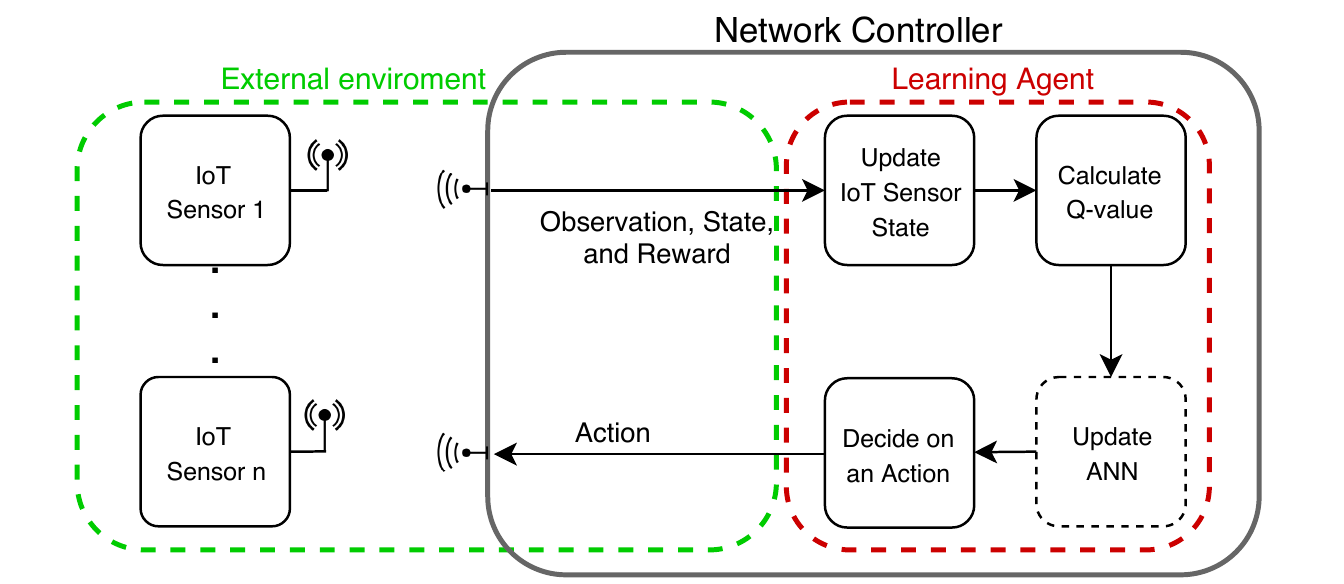}
	\caption{Diagram of the proposed updating mechanism.}
	\label{fig:learning_scheme}
	\vspace{-10pt}
\end{figure}

Figure \ref{fig:learning_scheme} shows a high-level model of our proposed mechanism. Sensors collecting information, along with the part of the network controller responsible for processing information, represent the external environment to our learning agent. An observation sent by an \ac{iot} sensor starts the learning cycle. The network controller then passes the necessary information (state and reward) to the learning agent. The learning agent then updates the state space and passes them to the \ac{ann}. The output of the \ac{ann} indicates which action the network controller should take (i.e., the action with the highest Q-value). The updating mechanism uses an $\epsilon$-greedy approach and; in our case, $\epsilon= 0.15$. Due to constant changes in the environment, the learning agent has to sometimes explore random actions to find the optimal action. In the last step, the network controller then transmits the action, i.e., the new update time, to the \ac{iot} sensor.

\begin{figure} 
	\centering
	\resizebox {0.95\columnwidth} {!} {
		\def\layersep{2.5cm}
\begin{tikzpicture}[shorten >=1pt,->,draw=black!50, node distance=\layersep]
    \tikzstyle{every pin edge}=[<-,shorten <=1pt]
    \tikzstyle{neuron}=[circle,fill=black!25,minimum size=17pt,inner sep=0pt]
    \tikzstyle{input neuron}=[neuron, fill=green!50];
    \tikzstyle{output neuron}=[neuron, fill=red!50];
    \tikzstyle{hidden neuron}=[neuron, fill=blue!50];
    \tikzstyle{annot} = [text width=4em, text centered]

    \node[input neuron, pin=left:$T_1\,\overline{\varepsilon}_1\,E_{1}$] (I-1) at (0,-1) {};
    \node[input neuron, pin=left:$T_2\,\overline{\varepsilon}_2\,E_{2}$] (I-2) at (0,-2) {};
    
    \node[input neuron, pin=left:$T_N\,\overline{\varepsilon}_N\,E_{N}$] (I-5) at (0,-5) {};

    \foreach \name / \y in {1,...,3}
        \path[yshift=0.5cm]
            node[hidden neuron] (H1-\name) at (\layersep,-\y ) {};
    
    \foreach \name / \y in {5,...,6}
        \path[yshift=0.5cm]
            node[hidden neuron] (H1-\name) at (\layersep,-\y ) {};

    \foreach \source in {1,...,2}
        \foreach \dest in {1,...,3}
            \path (I-\source) edge (H1-\dest);
    \foreach \source in {1,...,2}
        \foreach \dest in {5,...,6}
            \path (I-\source) edge (H1-\dest);

    \foreach \source in {5,...,5}
        \foreach \dest in {1,...,3}
            \path (I-\source) edge (H1-\dest);
            
    \foreach \source in {5,...,5}
        \foreach \dest in {5,...,6}
            \path (I-\source) edge (H1-\dest);
    
    \foreach \name / \y in {1,...,3}
        \path[yshift=0.5cm]
            node[hidden neuron] (H2-\name) at (2*\layersep,-\y ) {};
     
     \foreach \name / \y in {5,...,6}
        \path[yshift=0.5cm]
            node[hidden neuron] (H2-\name) at (2*\layersep,-\y ) {};
     \foreach \source in {1,...,3}
        \foreach \dest in {1,...,3}
            \path (H1-\source) edge (H2-\dest);
    \foreach \source in {1,...,3}
        \foreach \dest in {5,...,6}
            \path (H1-\source) edge (H2-\dest);
    
    \foreach \source in {5,...,6}
        \foreach \dest in {1,...,3}
            \path (H1-\source) edge (H2-\dest);
    \foreach \source in {5,...,6}
        \foreach \dest in {5,...,6}
            \path (H1-\source) edge (H2-\dest);
    
    \foreach \name / \y in {1,...,3}
        \path[yshift=0.5cm]
            node[hidden neuron] (H3-\name) at (3*\layersep,-\y ) {};
     
     \foreach \name / \y in {5,...,6}
        \path[yshift=0.5cm]
            node[hidden neuron] (H3-\name) at (3*\layersep,-\y ) {};
    
     \foreach \source in {1,...,3}
        \foreach \dest in {1,...,3}
            \path (H2-\source) edge (H3-\dest);
    \foreach \source in {1,...,3}
        \foreach \dest in {5,...,6}
            \path (H2-\source) edge (H3-\dest);
    
    \foreach \source in {5,...,6}
        \foreach \dest in {1,...,3}
            \path (H2-\source) edge (H3-\dest);
    \foreach \source in {5,...,6}
        \foreach \dest in {5,...,6}
            \path (H2-\source) edge (H3-\dest);
    
    \path (I-2) -- (I-5) node [black, font=\Huge, midway, sloped] {$\dots$};
    \path (H1-3) -- (H1-5) node [black, font=\Huge, midway, sloped] {$\dots$};
    \path (H2-3) -- (H2-5) node [black, font=\Huge, midway, sloped] {$\dots$};
    \path (H3-3) -- (H3-5) node [black, font=\Huge, midway, sloped] {$\dots$};

    
    \node[output neuron, pin={[pin edge={->}]right:$Q_{dec10}$}] (O-1) at (4*\layersep, -0.5 - 0.5) {};
    \node[output neuron, pin={[pin edge={->}]right:$Q_{dec1}$}] (O-2) at (4*\layersep,-1.5 - 0.5) {};
    \node[output neuron, pin={[pin edge={->}]right:$Q_{cons}$}] (O-3) at (4*\layersep,-2.5 - 0.5) {};
    \node[output neuron, pin={[pin edge={->}]right:$Q_{incr1}$}] (O-4) at (4*\layersep,-3.5- 0.5) {};
     \node[output neuron, pin={[pin edge={->}]right:$Q_{incr10}$}] (O-5) at (4*\layersep,-4.5- 0.5) {};
     
    
    \foreach \source in {1,...,3}
        \foreach \dest in {1,...,5}
            \path (H3-\source) edge (O-\dest);
    \foreach \source in {5,...,6}
        \foreach \dest in {1,...,5}
            \path (H3-\source) edge (O-\dest);
            


    \node[annot,above of=H2-1, node distance=1cm] (hl2) {Hidden layer 2};
    \node[annot,left of=hl2] (hl1){Hidden layer 1};
    \node[annot,left of=hl1] {Input layer };
    \node[annot,right of=hl2] (hl3){Hidden layer 3};
    \node[annot,right of=hl3] {Output layer};
\end{tikzpicture}
	}
	\caption{Neural network layout.}
	\label{fig:ann}
	\vspace{-10pt}
\end{figure}
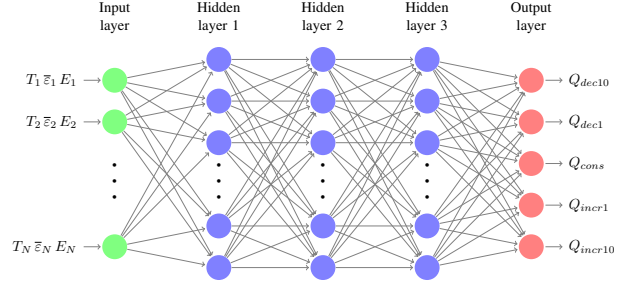


We implemented the \ac{ann} using PyTorch \cite{paszke2017automatic}, a deep learning Python library. Our \ac{ann} consists of three hidden layers, each with $1500$ neurons as presented in Fig. \ref{fig:ann}. The high number of neurons is required due to a possibly high number of inputs ($3N$). To train the \ac{ann}, the learning agent requires the values of state spaces of $N$ sensors and corresponding Q-values. The first inserted state space values ($T_1\,\overline{\varepsilon}_1\,E_{1}$) are from the sensor that transmitted last, followed by the state space values of the second last sensor ($T_2\,\overline{\varepsilon}_2\,E_{2}$), 
and so on.
We obtain the training Q-values by calculating the Q-value following Eq. \eqref{eq:q_value} for all five actions. We train the \ac{ann} only periodically, based on the latest observed behaviour, to shorten the response time.


Our learning agent is capable of responding  within $2-3$ $ms$, thus satisfying the timing constraints set by sensors' communications technology. For example, a device using \ac{lorawan} radio typically enters hibernation mode for more than a second. Additionally, response messages impose no additional energy cost to a sensor because the communication standard requires it to always listen to the channel before entering sleep mode.   

We evaluate our mechanism using data obtained from a real deployment, presented in the next section.


\section{Evaluation}
\label{sec:eval}
In this section, we evaluate our proposed solution using data provided by the Intel Berkeley Research laboratory \cite{bodik2004intel}. In our simulated network, sensor locations and transmitted observations (temperature and humidity) are based on the provided Intel data. We use nine days of measurements collected from 50 sensors. We list the static simulation parameters in Table \ref{table1}. The selected energy parameters mimic power consumption of an \ac{iot} sensor using \ac{lorawan} radio. We obtained the power parameters following the analysis presented in \cite{costa2017energy}.


In our simulations,  we set the time-step ($ts$) to $10$s. Each sensor starts with the same update time. We selected the starting update time for temperature, i.e., $T_{0t}$, and humidity, i.e., $T_{0h}$, by analyzing the dataset. We determined that if sensors transmit updates every $809s$ the average difference between two consecutive observation will be  $ \pm 0.25$ \textdegree{}C, and if they transmit a humidity observation every $606s$ the average difference will be $0.35\%$. For a fair comparison, throughout our evaluation, the updating mechanism keeps the average accuracy of temperature estimations within $0.25$ \textdegree{}C of the real temperature and within $0.35\%$ of the real air humidity percentage. To reduce the amount of required computations for \ac{mse} and estimation of real value we limit the number of used observations. We take eight observation from sensors closest to the estimation location.

\begin{table}[ht]
	\centering
	\caption{Simulation Parameters}
	\label{table1}
	\begin{tabular}{p{1.5cm} p{1.5cm}|| p{1.5cm} p{1.5cm}}
		\toprule
		\multicolumn{1}{l}{}                  
		\begin{tabular}[c]{@{}c@{}} Parameter \end{tabular} & 
		\begin{tabular}[c]{@{}c@{}} Value \end{tabular} & 
		\begin{tabular}[c]{@{}c@{}} Parameter \end{tabular} & 
		\begin{tabular}[c]{@{}c@{}} Value \end{tabular} \\ 
		
		\midrule 
		
		\rowcolor{Gray}
		\begin{tabular}[c]{@{}c@{}}$\phi_{acc}$  \end{tabular} & 
		\begin{tabular}[c]{@{}c@{}} $0.6$ \end{tabular}& 
		\begin{tabular}[c]{@{}c@{}}$\phi_{en}$ \end{tabular} & 
		\begin{tabular}[c]{@{}c@{}} $0.4$\end{tabular} \\
		
		\begin{tabular}[c]{@{}c@{}} $\alpha$  \end{tabular} & 
		\begin{tabular}[c]{@{}c@{}} $0.9$ \% \end{tabular}  & 
		\begin{tabular}[c]{@{}c@{}} $\gamma$ \end{tabular} &
		\begin{tabular}[c]{@{}c@{}}  $0.2$ \end{tabular} \\
		
		\rowcolor{Gray}
		
		\begin{tabular}[c]{@{}c@{}} $T_{0t}$  \end{tabular} & 
		\begin{tabular}[c]{@{}c@{}} $81$ $ts$ \end{tabular} & 
		\begin{tabular}[c]{@{}c@{}}$T_{0h}$ \end{tabular}  & 
		\begin{tabular}[c]{@{}c@{}} $61$ $ts$  \end{tabular}  \\

		\begin{tabular}[c]{@{}c@{}} time-step ($ts$)\end{tabular}  &   
		\begin{tabular}[c]{@{}c@{}}$10s$ \end{tabular} & 
		\begin{tabular}[c]{@{}c@{}}$P_c$\end{tabular} & 
		\begin{tabular}[c]{@{}c@{}}$30uW$\end{tabular} \\
		
		\rowcolor{Gray}
		
		\begin{tabular}[c]{@{}c@{}} $E_0$ \end{tabular} & 
		\begin{tabular}[c]{@{}c@{}} $6696 J$ \end{tabular} & 
		\begin{tabular}[c]{@{}c@{}} $\mathbb{E}[E_{tr}]$  \end{tabular}  & 
		\begin{tabular}[c]{@{}c@{}}  $63.7mJ$ \end{tabular}  \\
		
	    \begin{tabular}[c]{@{}c@{}} $\epsilon$\end{tabular} & 
		\begin{tabular}[c]{@{}c@{}}  $0.15$ \end{tabular} & 
		\begin{tabular}[c]{@{}c@{}}  \end{tabular}  & 
		\begin{tabular}[c]{@{}c@{}}  \end{tabular}  \\	
		
		\bottomrule
	\end{tabular}
\end{table}

In Fig. \ref{fig:learning_graphs} we show the change of update time and $\overline{\varepsilon}$ over a number of episodes for two sensors in a system of eight sensors. We iterate over the dataset five times. Each number in Fig. \ref{fig:learning_graphs} represents an end of a dataset iteration. In  Fig. \ref{fig:learning_graphs} (a) we plot the update time over a number of episodes for a sensor with above average available energy (95\%), while in \ref{fig:learning_graphs} (b) we plot update times of a sensors with below average energy (50\%). As we show, our updating mechanism sets the update time of a sensor with less energy significantly higher in comparison to the update time of a sensor with more energy available. Our updating mechanism is trying to balance the energy levels among sensors by setting a uneven update time. Simultaneously, as we show in Fig. \ref{fig:learning_graphs} (c) and (d) the agent is keeping the \ac{mse} close to the set target, i.e., $\varepsilon_{tar}$.


\begin{figure}
	\centering
\begin{tikzpicture}
\begin{groupplot}[group style={group name=my_plots, group size=2 by 2,vertical sep= 1.25cm}, width=0.3\textwidth]
\nextgroupplot[
axis lines=middle,
xlabel=episodes,
ylabel= $T_1$(min),
xmax=1600,ymax=285,
ymin=0,
xmin=0,
title = (a) $T$ for \ac{iot} Sensor 1,
legend style={at={(0.6,0.35)}},
ytick ={60, 120, 180, 240},
yticklabels={10, 20, 30, 40},
]
\draw[black!60,dashed] (0,240) -- (155, 240);
\draw[black!60,dashed] (0,180) -- (155, 180);
\draw[black!60,dashed] (0,120) -- (155, 120);
\draw[black!60,dashed] (0,60) -- (155, 60);

\addplot[line width=1pt,solid,color=blue]%
table[x=step,y=update_rates,col sep=comma]{csv_data/sensor_id22.csv};

\addplot[line width=2pt,dashed,color=red]%
table[x=step,y=update1,col sep=comma]{csv_data/update_time.csv};

\node [fill, draw, circle, minimum width=3pt, inner sep=0pt, pin={[outer sep=2pt]90:1}] at (48, 139) {};
\node [fill, draw, circle, minimum width=3pt, inner sep=0pt, pin={[outer sep=2pt]90:2}] at (101, 154) {};
\node [fill, draw, circle, minimum width=3pt, inner sep=0pt, pin={[outer sep=2pt]90:3}] at (152, 160) {};


\addlegendentry{$T$};
\addlegendentry{$T_{avg}$};

\nextgroupplot[
axis lines=middle,
xlabel=episodes,
ylabel= $T_2$(min),
xmax=1600,ymax=285,
ymin=0,
xmin=0,
title = (b) $T$ for \ac{iot} Sensor 2,
legend style={at={(0.6,0.35)}},
ytick ={60, 120, 180, 240},
yticklabels={10, 20, 30, 40},
]
\draw[black!60,dashed] (0,240) -- (500, 240);
\draw[black!60,dashed] (0,180) -- (500, 180);
\draw[black!60,dashed] (0,120) -- (500, 120);
\draw[black!60,dashed] (0,60) -- (500, 60);
\addplot[line width=1pt,solid,color=blue]%
table[x=step,y=update_rates,col sep=comma]{csv_data/sensor_id23.csv};

\addplot[line width=2pt,dashed,color=red]%
table[x=step,y=update2,col sep=comma]{csv_data/update_time.csv};

\node [fill, draw, circle, minimum width=3pt, inner sep=0pt, pin={[outer sep=2pt]270:1}] at (39, 216) {};
\node [fill, draw, circle, minimum width=3pt, inner sep=0pt, pin={[outer sep=2pt]270:2}] at (76, 224) {};
\node [fill, draw, circle, minimum width=3pt, inner sep=0pt, pin={[outer sep=2pt]270:3}] at (113, 197) {};

%
\addlegendentry{$T$};
\addlegendentry{$T_{avg}$};

\nextgroupplot[
axis lines=middle,
xlabel=episodes,
ylabel= $\overline{\varepsilon}_1(\%)$,
xmax=1600,ymax=0.017,
ymin=0,
xmin=0,
title = (c) $\overline{\varepsilon}$ for IoT sensor 1,
legend style={at={(0.6,0.35)}},
ytick ={0.005,0.01, 0.0125, 0.015},
yticklabels={0.5,1, 1.25, 1.5},
ytick scale label code/.code={}
]
\draw[black!60,dashed] (0,200) -- (700, 200);
\draw[black!60,dashed] (0,150) -- (700, 150);
\draw[black!60,dashed] (0,100) -- (700, 100);
\draw[black!60,dashed] (0,50) -- (700, 50);

\addplot[line width=1pt,solid,color=blue]%
table[x=step,y=estimation_errors,col sep=comma]{csv_data/sensor_id22.csv};

\addplot[line width=1pt,dashed,color=red]%
table[x=step,y=error,col sep=comma]{csv_data/error_line_learn.csv};

\node [fill, draw, circle, minimum width=3pt, inner sep=0pt, pin={[outer sep=2pt]90:1}] at (48, 90) {};
\node [fill, draw, circle, minimum width=3pt, inner sep=0pt, pin={[outer sep=2pt]90:2}] at (101, 102) {};
\node [fill, draw, circle, minimum width=3pt, inner sep=0pt, pin={[outer sep=2pt]90:3}] at (152, 107) {};

\addlegendentry{$\overline{\varepsilon}$};
\addlegendentry{ $\varepsilon_{tar}$};

\nextgroupplot[
axis lines=middle,
xlabel=episodes,
ylabel= $\overline{\varepsilon}_2(\%)$,
xmax=1600,ymax=0.017,
ymin=0,
xmin=0,
title = (d) $\overline{\varepsilon}$ for IoT sensor 2,
legend style={at={(0.6,0.35)}},
ytick ={0.005,0.01, 0.0125, 0.015},
yticklabels={0.5,1, 1.25, 1.5},
ytick scale label code/.code={}
]
\draw[black!60,dashed] (0,200) -- (700, 200);
\draw[black!60,dashed] (0,150) -- (700, 150);
\draw[black!60,dashed] (0,100) -- (700, 100);
\draw[black!60,dashed] (0,50) -- (700, 50);

\addplot[line width=1pt,solid,color=blue]%
table[x=step,y=estimation_errors,col sep=comma]{csv_data/sensor_id23.csv};

\addplot[line width=1pt,dashed,color=red]%
table[x=step,y=error,col sep=comma]{csv_data/error_line_learn.csv};

\node [fill, draw, circle, minimum width=3pt, inner sep=0pt, pin={[outer sep=2pt]270:1}] at (39, 140) {};
\node [fill, draw, circle, minimum width=3pt, inner sep=0pt, pin={[outer sep=2pt]270:2}] at (76, 140) {};
\node [fill, draw, circle, minimum width=3pt, inner sep=0pt, pin={[outer sep=2pt]270:3}] at (113, 127) {};

\addlegendentry{$\overline{\varepsilon}$};
\addlegendentry{ $\varepsilon_{tar}$};

\end{groupplot}

\end{tikzpicture}
	\caption{Two IoT sensors learning  over a number of episodes. The numbers in the graphs mark the end of an iteration over the dataset.}
	\label{fig:learning_graphs}
	\vspace{-10pt}
\end{figure}
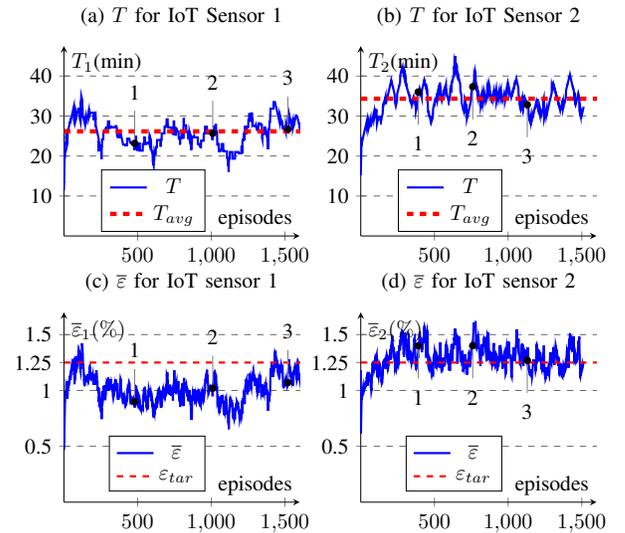


\begin{figure} 
	\centering
\begin{tikzpicture}
\begin{groupplot}[group style={group name=my_plots, group size=1 by 2,vertical sep= 1.75cm}, width=0.75\textwidth, height=0.3\textwidth,]
\nextgroupplot[
axis lines=middle,
xlabel=episodes,
ylabel= $T$(min),
xmax=208,ymax=250,
ymin=0,
xmin=0,
title = (a) Update time over a number of episodes,
legend style={at={(0.2655,0.3)}},
ytick ={30, 60, 90, 120, 150, 180, 210, 240},
yticklabels={5,10,15,20,25,30, 35, 40},
]

\draw[black!60,dashed] (0,180) -- (700, 180);
\draw[black!60,dashed] (0,120) -- (700, 120);
\draw[black!60,dashed] (0,60) -- (700, 60);

\draw[black!30,dashed] (200,0) -- (200, 500);
\draw[black!30,dashed] (350,0) -- (350, 500);

\coordinate [label=-90:new $\theta_1$  $\theta_2$]  (A) at (50,105)  ;
\coordinate  (B) at (50,145) ;
\draw [->,line width=1mm] (A) -- (B);

\coordinate [label=-90:new $\theta_1$  $\theta_2$]  (C) at (100, 80)  ;
\coordinate  (D) at (100,120) ;
\draw [->,line width=1mm] (C) -- (D);

\coordinate [label=-90:new $\varepsilon_{tar}$]  (E) at (150, 80)  ;
\coordinate  (F) at (150,120) ;
\draw [->,line width=1mm] (E) -- (F);

\addplot[line width=2pt,solid,color=blue]%
table[x=step,y=update_time,col sep=comma]{csv_data/optimal_temperature.csv};

\addplot[line width=2pt,dashed,color=red]%
table[x=step,y=optimal,col sep=comma]{csv_data/optimal_line.csv};

\addlegendentry{Learned $T$};
\addlegendentry{Optimal $T$};

\nextgroupplot[
axis lines=middle,
xlabel=episodes,
ylabel= $\overline{\varepsilon}$ ($\%$),
xmax=208,ymax=0.023,
ymin=0,
xmin=0,
title = (b) Average MSE over a number of episodes,
legend style={at={(0.21,0.25)}},
ytick ={0.005,0.01, 0.0125, 0.015, 0.02},
yticklabels={0.5,1, 1.25, 1.5, 2},
ytick scale label code/.code={}
]


\draw[black!60,dashed] (0,200) -- (700, 200);
\draw[black!60,dashed] (0,150) -- (700, 150);
\draw[black!60,dashed] (0,100) -- (700, 100);
\draw[black!60,dashed] (0,50) -- (700, 50);

\draw[black!30,dashed] (200,0) -- (200, 500);
\draw[black!30,dashed] (350,0) -- (350, 500);

\coordinate [label=-270:new $\theta_1$  $\theta_2$]  (A) at (50,170)  ;
\coordinate  (B) at (50,130) ;
\draw [->,line width=1mm] (A) -- (B);

\coordinate [label=-90:new $\theta_1$  $\theta_2$]  (C) at (100, 80)  ;
\coordinate  (D) at (100,120) ;
\draw [->,line width=1mm] (C) -- (D);

\coordinate [label=-90:new $\varepsilon_{tar}$]  (E) at (150, 80)  ;
\coordinate  (F) at (150,120) ;
\draw [->,line width=1mm] (E) -- (F);

\addplot[line width=2pt,solid,color=blue]%
table[x=step,y=error,col sep=comma]{csv_data/optimal_temperature.csv};
\addplot[line width=2pt,dashed,color=red]%
table[x=step,y=error,col sep=comma]{csv_data/error_line.csv};

%
\addlegendentry{$\varepsilon$};
\addlegendentry{ $\varepsilon_{tar}$};
\end{groupplot}

\end{tikzpicture}
	\caption{Updating mechanism searching for the optimal update time, arrows indicate change in the covariance model scaling parameters or $\varepsilon_{tar}$.}
	\label{fig:optimal_behaviour}
\end{figure}
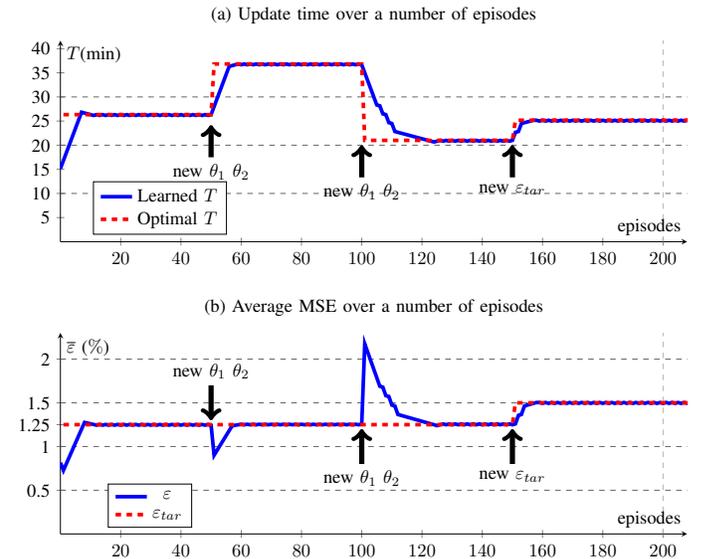

To show that the updating mechanism is capable of finding the optimal solution, i.e., capable of determining the maximal update time possible, we test the mechanism's performance in a system with only two sensors. We set only one sensor in the learning mode while the other keeps the update time constant. Furthermore, covariance model scaling parameters changed only in selected episodes. Such a system enables us to also obtain the optimal solution for comparison purposes. We changed the scaling parameters at episode $50$ and $100$. At episode $150$ we changed $\varepsilon_{tar}$. The mechanism is always capable of adapting and finding the optimal update time, as shown in  Fig. \ref{fig:optimal_behaviour} (a). In Fig. \ref{fig:optimal_behaviour} (b), we show the change of $\overline{\varepsilon}$ over a number of episodes: the mechanism always converges toward the selected $\varepsilon_{tar}$.

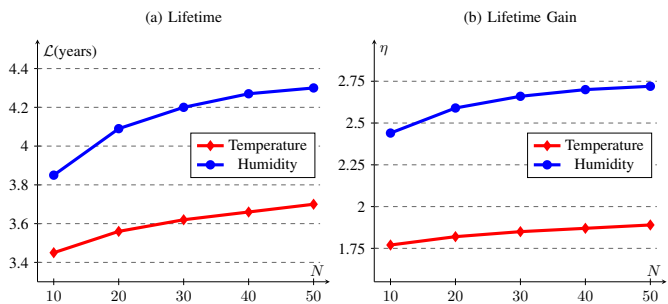
\begin{figure}
	\centering

\begin{tikzpicture}
\begin{groupplot}[group style={group name=my_plots, group size=2 by 1,vertical sep= 1.75cm}, width=0.45\textwidth]]





\nextgroupplot[
axis lines=middle,
title = (a) Lifetime,
xlabel=$N$,
ylabel=$\mathcal{L}$(years),
ytick ={3.4, 3.6, 3.8, 4, 4.2, 4.4},
yticklabels={3.4, 3.6, 3.8, 4, 4.2, 4.4},,
xtick={10,20,30,40,50},
legend style={at={(0.95,0.615)}},
xmin= 7.5,xmax=52.5,
ymin= 3.3,ymax= 4.55,
every tick/.style={
	black,
	semithick,
}]
\draw[black!60,dashed] (0,110) -- (435, 110);
\draw[black!60,dashed] (0,90) -- (435, 90);
\draw[black!60,dashed] (0,70) -- (435, 70);
\draw[black!60,dashed] (0,50) -- (435, 50);
\draw[black!60,dashed] (0,30) -- (435, 30);
\draw[black!60,dashed] (0,10) -- (435, 10);

\addplot[red, line width=2pt, mark=diamond*] 
table[x=density,y=tem_life,col sep=comma]{csv_data/our_approach.csv};

\addplot[blue, line width=2pt, mark=*] %
table[x=density,y=hum_life,col sep=comma]{csv_data/our_approach.csv};

\addlegendentry{Temperature};
\addlegendentry{Humidity};




\nextgroupplot[
axis lines=middle,
title = (b) Lifetime Gain,
xlabel=$N$,
ylabel= $\eta$,
ytick ={1.75,2,2.25, 2.5, 2.75},
yticklabels={1.75,2,2.25, 2.5, 2.75},,
xtick={10,20,30,40,50},
legend style={at={(0.95,0.615)}},
xmin= 7.5,xmax=52.5,
ymin= 1.55,ymax= 3,
every tick/.style={
	black,
	semithick,
}]
\draw[black!60,dashed] (0,120) -- (435, 120);
\draw[black!60,dashed] (0,95) -- (435, 95);
\draw[black!60,dashed] (0,70) -- (435, 70);
\draw[black!60,dashed] (0,45) -- (435, 45);
\draw[black!60,dashed] (0,20) -- (435, 20);

\addplot[red, line width=2pt, mark=diamond*] 
table[x=density,y=tem_relative_incr,col sep=comma]{csv_data/our_approach.csv};

\addplot[blue, line width=2pt, mark=*] %
table[x=density,y=hum_relative_incr,col sep=comma]{csv_data/our_approach.csv};

\addlegendentry{Temperature};
\addlegendentry{Humidity};

\end{groupplot}
\end{tikzpicture}
	\caption{Sensors' lifetime and lifetime gain achieved by our updating mechanism as the number of sensors under its control increases.}
	\label{fig:energy_savings}
	\vspace{-10pt}
\end{figure}

Next, we test updating mechanism performance as the number of sensors, $N$, under its management increases. The expected lifetime of sensors increases with more sensors in the network, as we show in Fig. \ref{fig:energy_savings} (a).  The gain for using correlated information is higher when observing humidity, due to higher correlation exhibited in the observations collected. We calculate the expected lifetime using Eq. \eqref{eq:lifetime}. In our calculation, we use average update time our updating mechanism achieves on the $9^{th}$ day. We assume that each sensor is equipped with a $620 mAh$ Lithium battery. To show the improvement in comparison to the baseline case, we calculate the lifetime gain, i.e., $\eta_i $, as the ratio between the lifetime achieved using our mechanism and in the baseline case:

\begin{equation}\label{eq:lifetime_ratio}
\eta = \frac{\mathbb{E}[ \mathcal{L} ] }{\mathbb{E}[ \mathcal{L}_0 ] } .
\end{equation}

\noindent We calculated the baseline $\mathbb{E}[ \mathcal{L}_0 ]$ using $T_{0t}$ and $T_{0h}$, resulting in the expected lifetime of $1.95$ years when observing temperature, and $1.56$ years when observing humidity. Our updating mechanism can significantly prolong the expected lifetime of battery-powered devices. When measuring humidity the expected sensor lifetime can be extended to over four years, while for sensors measuring temperature the expected lifetime extends to over three and a half years.

\section{Final Remarks and Future Work}
\label{sec:conclusion}
In this paper, we applied deep Q-learning to prolong the lifetime of battery-powered sensors. The proposed updating mechanism is capable of adjusting updates according to a sensor's available energy and the sensed accuracy of the observed physical phenomenon. We demonstrated that it is capable of performing in a dynamic \ac{iot} network environment. We evaluated our proposed mechanism using data obtained from a real deployment. Our results show that it is possible to increase the lifetime of battery-powered sensors by a factor of three by taking advantage of correlated information.

 In our future work, we will consider a network of sensors using different primary power sources, e.g., mains powered or event-based harvesting. To provide the network controller with a capability to manage such devices effectively, we will expand the list of available actions. Additionally, we will design a new reward function to reflect the different energy sources across the sensors. In such a network, the primary goal of learning will be achieving an energy-aware balancing scheduling of sensors' updates.






\section*{Acknowledgments}
This work was funded in part by the European Regional Development Fund through the SFI Research Centres Programme under Grant No. 13/RC/2077 SFI CONNECT and by the SFI-NSFC Partnership Programme Grant Number 17/NSFC/5224.


\begin{acronym}[MACHU]
  \acro{cr}[CR]{Cognitive Radio}
  \acro{ofdm}[OFDM]{orthogonal frequency-division multiplexing}
  \acro{ofdma}[OFDMA]{orthogonal frequency-division multiple access}
  \acro{scfdma}[SC-FDMA]{single carrier frequency division multiple access}
  \acro{rbi}[RBI]{ Research Brazil Ireland}
  \acro{rfic}[RFIC]{radio frequency integrated circuit}
  \acro{sdr}[SDR]{Software Defined Radio}
  \acro{sdn}[SDN]{Software Defined Networking}
  \acro{su}[SU]{Secondary User}
  \acro{ra}[RA]{Resource Allocation}
  \acro{qos}[QoS]{quality of service}
  \acro{usrp}[USRP]{Universal Software Radio Peripheral}
  \acro{mno}[MNO]{Mobile Network Operator}
  \acro{mnos}[MNOs]{Mobile Network Operators}
  \acro{gsm}[GSM]{Global System for Mobile communications}
  \acro{tdma}[TDMA]{Time-Division Multiple Access}
  \acro{fdma}[FDMA]{Frequency-Division Multiple Access}
  \acro{gprs}[GPRS]{General Packet Radio Service}
  \acro{msc}[MSC]{Mobile Switching Centre}
  \acro{bsc}[BSC]{Base Station Controller}
  \acro{umts}[UMTS]{universal mobile telecommunications system}
  \acro{Wcdma}[WCDMA]{Wide-band code division multiple access}
  \acro{wcdma}[WCDMA]{wide-band code division multiple access}
  \acro{cdma}[CDMA]{code division multiple access}
  \acro{lte}[LTE]{Long Term Evolution}
  \acro{papr}[PAPR]{peak-to-average power rating}
  \acro{hn}[HetNet]{heterogeneous networks}
  \acro{phy}[PHY]{physical layer}
  \acro{mac}[MAC]{medium access control}
  \acro{amc}[AMC]{adaptive modulation and coding}
  \acro{mimo}[MIMO]{multiple input multiple output}
  \acro{rats}[RATs]{radio access technologies}
  \acro{vni}[VNI]{visual networking index}
  \acro{rbs}[RB]{resource blocks}
  \acro{rb}[RB]{resource block}
  \acro{ue}[UE]{user equipment}
  \acro{cqi}[CQI]{Channel Quality Indicator}
  \acro{hd}[HD]{half-duplex}
  \acro{fd}[FD]{full-duplex}
  \acro{sic}[SIC]{self-interference cancellation}
  \acro{si}[SI]{self-interference}
  \acro{bs}[BS]{base station}
  \acro{fbmc}[FBMC]{Filter Bank Multi-Carrier}
  \acro{ufmc}[UFMC]{Universal Filtered Multi-Carrier}
  \acro{scm}[SCM]{Single Carrier Modulation}
  \acro{isi}[ISI]{inter-symbol interference}
  \acro{ftn}[FTN]{Faster-Than-Nyquist}
  \acro{m2m}[M2M]{machine-to-machine}
  \acro{mtc}[MTC]{machine type communication}
  \acro{mmw}[mmWave]{millimeter wave}
  \acro{bf}[BF]{beamforming}
  \acro{los}[LOS]{line-of-sight}
  \acro{nlos}[NLOS]{non line-of-sight}
  \acro{capex}[CAPEX]{capital expenditure}
  \acro{opex}[OPEX]{operational expenditure}
  \acro{ict}[ICT]{information and communications technology}
  \acro{sp}[SP]{service providers}
  \acro{inp}[InP]{infrastructure providers}
  \acro{mvnp}[MVNP]{mobile virtual network provider}
  \acro{mvno}[MVNO]{mobile virtual network operator}
  \acro{nfv}[NFV]{network function virtualization}
  \acro{vnfs}[VNF]{virtual network functions}
  \acro{cran}[C-RAN]{Cloud Radio Access Network}
  \acro{bbu}[BBU]{baseband unit}
  \acro{bbus}[BBU]{baseband units}
  \acro{rrh}[RRH]{remote radio head}
  \acro{rrhs}[RRH]{Remote radio heads} 
  \acro{sfv}[SFV]{sensor function virtualization}
  \acro{wsn}[WSN]{wireless sensor networks} 
  \acro{bio}[BIO]{Bristol is open}
  \acro{vitro}[VITRO]{Virtualized dIstributed plaTfoRms of smart Objects}
  \acro{os}[OS]{operating system}
  \acro{www}[WWW]{world wide web}
  \acro{iotvn}[IoT-VN]{IoT virtual network}
  \acro{mems}[MEMS]{micro electro mechanical system}
  \acro{mec}[MEC]{Mobile edge computing}
  \acro{coap}[CoAP]{Constrained Application Protocol}
  \acro{vsn}[VSN]{Virtual sensor network}
  \acro{rest}[REST]{REpresentational State Transfer}
  \acro{aoi}[AoI]{Age of Information}
  \acro{lora}[LoRa\texttrademark]{Long Range}
  \acro{iot}[IoT]{Internet of Things}
  \acro{snr}[SNR]{Signal-to-Noise Ratio}
  \acro{cps}[CPS]{Cyber-Physical System}
  \acro{uav}[UAV]{Unmanned Aerial Vehicle}
  \acro{rfid}[RFID]{Radio-frequency identification}
  \acro{lpwan}[LPWAN]{Low-Power Wide-Area Network}
  \acro{lgfs}[LGFS]{Last Generated First Served}
  \acro{wsn}[WSN]{wireless sensor network} 
  \acro{lmmse}[LMMSE]{Linear Minimum Mean Square Error}
  \acro{rl}[RL]{Reinforcement Learning}
  \acro{nb-iot}[NB-IoT]{Narrowband IoT}
  \acro{lorawan}[LoRaWAN]{Long Range Wide Area Network}
  \acro{mdp}[MDP]{Markov Decision Process}
  \acro{ann}[ANN]{Artificial Neural Network}
  \acro{dqn}[DQN]{Deep Q-Network}
  \acro{mse}[MSE]{Mean Square Error}
\end{acronym}

\bibliographystyle{templates/IEEEtran}  
\bibliography{paper.bib}

\end{document}